\documentclass[twocolumn,aps,superscriptaddress,longbibliography]{revtex4-2}
\usepackage{comment}
\usepackage{amssymb}
\usepackage{amsfonts}
\usepackage{amsmath}
\usepackage{amsxtra}
\usepackage{amscd}
\usepackage{bm}
\usepackage{amsthm}
\usepackage{graphicx}
\usepackage{epsf}
\usepackage{bbold}
\usepackage{xcolor}
\usepackage{mathalfa}
\usepackage{makecell} 
\usepackage{multirow}
\usepackage{hhline}
\usepackage{tikz}
\usepackage{physics}
\usepackage{eucal}
\usepackage{stackengine}
\usepackage{feynmp-auto}

\newcommand{\beq}{\begin{equation}}
\newcommand{\eeq}{\end{equation}}
\newcommand{\bmul}{\begin{multline}}
\newcommand{\emul}{{\end{multline}}}
\newcommand\beqa{\begin{eqnarray}}
\newcommand\eeqa{\end{eqnarray}}
\newcommand\bea{\begin{array}}
\newcommand\eea{\end{array}}
\newcommand\ba{\begin{array}}
\newcommand\ea{\end{array}}

\newcommand{\neqa}{\nonumber\end{eqnarray}}

\usepackage[export]{adjustbox}

\makeatletter
\protected\def\xvcenter{%
  \hbox\bgroup$\everyvbox{\everyvbox{}\aftergroup\m@th\aftergroup$\aftergroup\egroup}%
  \vcenter
}

\DeclareRobustCommand{\midscript}[1]{
  \mathchoice{\mid@script\scriptstyle{#1}}
    {\mid@script\scriptstyle{#1}}
    {\mid@script\scriptscriptstyle{#1}}
    {\mid@script\scriptscriptstyle{#1}}
}
\newcommand{\mid@script}[2]{
  \vcenter{\hbox{$\m@th#1#2$}}
}

\begin{document}

\title{Up/down-conversion of infrared light by few-layer graphene polytypes}

\author{Patrick Johansen Sarsfield}
\affiliation{National Graphene Institute, University of Manchester, Manchester M13 9PL, United Kingdom}
\affiliation{Department of Physics \& Astronomy, University of Manchester, Manchester M13 9PL, United Kingdom}

\author{Takaaki V. Joya}
\affiliation{Department of Physics, University of Osaka, Toyonaka, Osaka 560-0043, Japan}

\author{Takuto Kawakami}
\affiliation{Center for Integrated Science and Humanities, Fukushima Medical University, Fukushima 960-1295, Japan}

\author{Mikito Koshino}
\affiliation{Institute for Solid State Physics, The University of Tokyo, Kashiwa, Chiba 277-8581, Japan}

\author{Vladimir Fal'ko}
\affiliation{National Graphene Institute, University of Manchester, Manchester M13 9PL, United Kingdom}
\affiliation{Department of Physics \& Astronomy, University of Manchester, Manchester M13 9PL, United Kingdom}

\begin{abstract}
Optical nonlinearity of materials with broken inversion symmetry enables two-photon processes where an incoming pair of photons can generate an up-converted photon with the combined frequency, or a high-energy photon can be split into a correlated pair of down-converted photons. Here, we identify few-layer graphene films that offer up/down-conversion capability in the infrared spectral range. For mixed stacking tetralayers (ABCB), which are non-centrosymmetric 2D crystals, we find highly efficient nearly resonant up/down-conversion of [$\omega_1,\omega_2$] photon pairs into/from a photon with $0.7$eV$<\Omega=\omega_1+\omega_2<1.1$eV. We also note that a pronounced second-order nonlinearity in the spectral range of $0.7$eV$<\Omega<0.9$eV can be promoted in rhombohedral tri- and tetralayers by asymmetrical encapsulation. Potentially, for fibers coated with few-layer graphene, this opens the door for ``in fiber" production of photon pairs with correlated polarizations.

\end{abstract}

\maketitle

\section{Introduction}
Second-order non-linear optical phenomena can arise when a beam of light illuminates a material in the absence of inversion symmetry \cite{RosencherBoisNagle1996}. This phenomena has been studied extensively as a probe of structural and electronic symmetry in a wide range of materials \cite{SchanklerRappe2021,TrilayerSHG,vandelli2019resonant,Qian2023,Jiang2021}. For monochromatic illumination, one observes either the bulk photovoltaic effect or second-harmonic generation (SHG); when two distinct frequencies of light are incident on a material, the analogous response is up-conversion also known as sum-frequency generation (SFG)~\cite{busson2023sum}, a natural generalization of SHG to non-degenerate frequencies.
Simultaneously, one expects to see the reverse process of spontaneous down conversion (SDC, also known as parametric down-conversion ~\cite{Zhang2021SPDCReview,SPDCFilms2020}), in which a single higher-energy photon splits into a correlated pair of lower-energy photons subject to energy conservation.
Down conversion is of particular importance in quantum information science, where it is the workhorse process for generating entangled photon pairs for use in quantum cryptography, quantum metrology, and tests of quantum nonlocality~\cite{Zhang2021SPDCReview,SPDCFilms2020}.

Graphene and its multilayers offer an appealing, electrically tunable platform for engineering such second-order responses.
Although monolayer graphene's centrosymmetry forbids a dipole-allowed second-order nonlinearity, this symmetry is broken in some Bernal- and all mixed-stacking multilayers and can be further controlled by an applied displacement field or substrate-induced asymmetry which can break the inversion symmetry in all graphene multilayers.
In fact, SHG has been used experimentally to probe stacking-order symmetry breaking in bilayer, trilayer, and tetralayer graphene~\cite{vandelli2019resonant,Zhang2022SHG,TrilayerSHG,ZhouZhuXu2024}.
The electronic structure of multilayer graphene depends sensitively on the number of layers and the stacking sequence, giving an additional, gate- and growth-controlled degree of freedom absent in conventional nonlinear crystals~\cite{KoshinoMcCann2013,KoshinoMcCann2009,McCannKoshino2013RPP}.
While the resulting SHG response has been mapped for several stacking orders~\cite{BrunPedersen2015,sarsfield2026resonant}, its extension to general, non-degenerate frequency combinations -- and in particular its implications for down-conversion and entangled-photon-pair generation -- has not been systematically explored.
Here we compute the frequency-resolved second-order optical conductivity of bilayer, trilayer, and tetralayer graphene for all stacking orders, and show that stacking order and electrostatic gating together provide a versatile means of controlling the double-resonance features responsible for both up and down conversion, including narrow-band down-conversion into telecom-relevant photon-pair frequencies suitable for entangled-photon generation.

\begin{comment}
Non-linear optical phenomena can occur in cases where a strong beam of light is shone onto a material, or inversion symmetry is broken.
In the former, one would see odd-high harmonic generation, while the latter gives rise to even-orders.
The second-order non-linearity is an optical phenomenon that has been studied widely.
In cases where monochromatic light is illuminated, one would see either the bulk photovoltaic effect or the second-harmonic generation.
On the other hand, one could imagine a situation where two frequencies of light are shone onto a material.
In this case, we expect to see an analogous response to the SHG, which is referred to as sum-frequency generation (SFG).
Simultaneously, one can expect to see its reverse process, known as spontaneous parametric down-conversion (SPDC).
\end{comment}

\section{Methods}
\begin{figure*}[hbt!]
\centering
  { \includegraphics[width=1\textwidth]{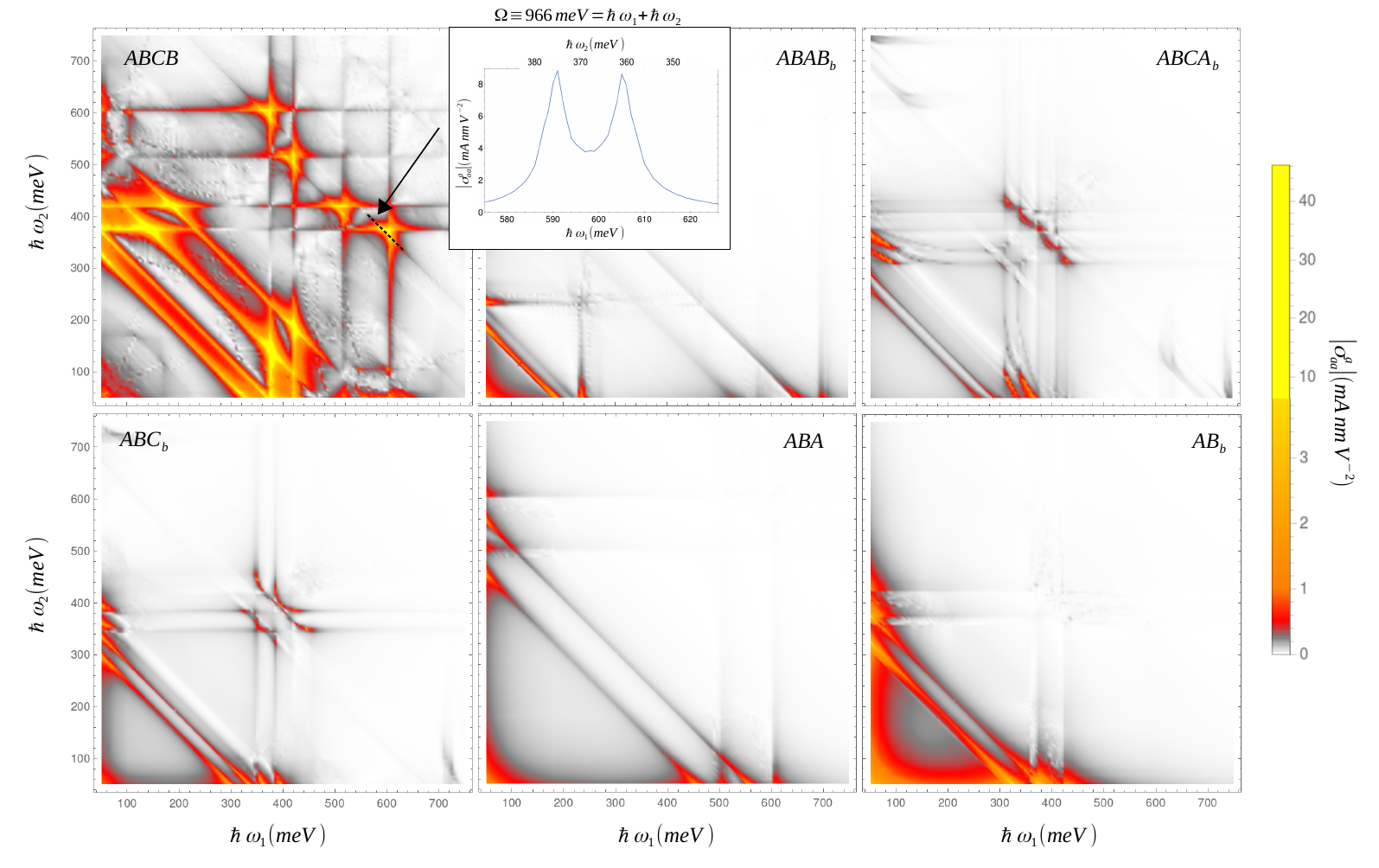}}
\caption{\label{fig:2}
The absolute value of second order conductivity for all stacking orders up to tetralayer. The inversion symmetric stacking orders have their symmetry broken by an on-layer proximity induced potential representing a substrate. The axis on each panel are the energies of the two lower energy photons respectively. For the inset we analise an exemplary resonance of ABCB graphene. Here we show how one may use the provided calculations to predict down conversion rates. For this example, the dashed line represents the $\Omega\equiv966$meV$=\hbar\omega_1+\hbar\omega_2$ line. The inset graph follows this line cut showing modulus of conductivity as a function of the two photon energies (on the upper and lower horizontal axis) that are produced during down conversion. The square of this conductivity is a probability distribution for the resultant pair produced photons. 
}
\end{figure*}

The second-order optical nonlinearity is a process in which three photons are involved.
Here, we compute the interacting Hamiltonian between the electron and the photons.

We begin by considering the total Hamiltonian $H$ of the system, including both the electron and the quantized photon field, given as
\begin{equation}
    H=H_{\rm el}+H_{\rm ph}+H_{\rm int},
\end{equation}
where $H_{\rm el}$, $H_{\rm ph}$, and $H_{\rm int}$ are the electronic, photonic, and interaction Hamiltonian, respectively.
The electronic Hamiltonian is the $\vb{k}\cdot\vb{p}$ tight-binding Hamiltonian, given explicitly in our previous works Ref~\cite{sarsfield2024substrate,sarsfield2026resonant}.

The multilayer Hamiltonian will be implemented for two types of systems: (i) Bernal stacking ABA trilayers and mixed stacking ABCB tetralayers whose crystalline lattice lacks inversion symmetry, and (ii) bilayers and rhombohedral multilayers where we permit an encapsulation induced inversion symmetry breaking. In the latter case, we account for a proximity-induced on-layer potential, $\Delta$, on the bottom layer (indicated by subscript "b" in Fig. \ref{fig:2}) that touches a substrate (in the quantitative analysis we use $\Delta=18$meV indicative of an hBN substrate \cite{boschi_built-bernal_2024}). For illustration purposes, we also analyse in more detail the effect of externally induced asymmetry in bilayers where an encapsulation-induced on-layer energy difference, $\Delta$, between the top and bottom layers may be enhanced by electrostatic gating (in that case we put on-layer energy $\pm\Delta/2$ on the top/bottom layer).

The photons and their fields are accounted by 
\begin{align}
     H_{\rm ph}&=\sum_{\omega,\mu}\hbar\omega\,a^\dagger_{\omega,\mu} a_{\omega,\mu},\nonumber
    \\
    {\bf A}(t)&=\sum_{\omega,\mu}\sqrt{\frac{\hbar}{2\varepsilon_0\omega V}}\left({\bf e}_\omega\,a_{\omega,\mu}\,e^{-i\omega t}+{\bf e}^*_\omega\,a^\dagger_{\omega,\mu}\,e^{i\omega t}\right),
    \label{eq:1}
\end{align}

where $a^\dagger_{\omega,\mu}$ and $a_{\omega,\mu}$ are the creation and annihilation operators of a photon with frequency $\omega$ with a unit polarization vector $\vb{e}_\omega$ in the $\mu$ direction, $V$ is the quantitation volume, and $\varepsilon_0$ is the vacuum permittivity. The interaction between the electrons and the photons is given by
\begin{align}
    H_{\rm int}&=-\int d^3r\sum_\mu j^\mu(t)\,A^\mu(t) \nonumber
    \\
    &=-V\sum_\mu j^\mu(t)\,A^\mu(t).
    \label{eq:minimalcoupling}
\end{align}

From this, the two photon processes can be described using an effective nonlinear coupling,  

\begin{widetext}
\begin{align}
    H_{\rm int}&=-\frac{1}{\sqrt{V}}\left(\frac{\hbar}{2\varepsilon_0}\right)^{\frac{3}{2}}\frac{1}{\sqrt{\omega_1\omega_2\Omega}}\sum_{\mu,\alpha,\beta}\left\{\Pi^\mu_{\alpha\beta}\,(e^\mu_\Omega)^*e^\alpha_{\omega_1}e^\beta_{\omega_2}\,a_{\omega_1,\alpha}\,a_{\omega_2,\beta}\,a^\dagger_{\Omega,\mu}+\left[\Pi^\mu_{\alpha\beta}\,(e^\mu_\Omega)^*e^\alpha_{\omega_1}e^\beta_{\omega_2}\right]^*a^\dagger_{\omega_1,\alpha}\,a^\dagger_{\omega_2,\beta}\,a_{\Omega,\mu}\right\},\nonumber
    \\
    &=-\frac{1}{\sqrt{V}}\left(\frac{\hbar}{2\varepsilon_0}\right)^{\frac{3}{2}}\frac{1}{\sqrt{\omega_1\omega_2\Omega}}\sum_{\mu,\alpha,\beta}\left(
    \vcenter{\hbox{
        \begin{fmffile}{upconversion}
        \begin{fmfgraph*}(50,60)
            \fmfcmd{
                path q; 
                q = (0,0) -- (0.5,0) & quartercircle & (0,0.5) -- (0,0);
                def draw_otimes(expr p) = 
                    filldraw (for i=0 upto 3: q rotated (45+90*i) shifted point 0 of p -- endfor cycle) withpen pencircle scaled 0.4;
                enddef;
            }
            \fmfleft{p1,p2,i1,q1,i2,p3,p4}
            \fmfright{o1}
            \fmfforce{(0.1w,0.2h)}{i1}
            \fmfforce{(0.1w,0.8h)}{i2}
            \fmfforce{(0.85w,0.5h)}{o1}
            \fmfforce{(0.4w,0.5h)}{v1}
            \fmf{photon, arrow, arrow.size=8pt}{v1,o1}
            \fmflabel{$\mu$}{o1}
            \fmf{photon, arrow, arrow.size=8pt}{i1,v1}
            \fmflabel{$\beta$}{i1}
            \fmf{photon, arrow, arrow.size=8pt}{i2,v1}
            \fmflabel{$\alpha$}{i2}
            \fmfv{decor.shape=circle, decor.filled=shaded, decor.size=0.16h}{v1}
            %\fmfdot{v1}
            %\fmfv{decor.shape=otimes,decor.filled=empty,decor.size=1h}{v1}
        \end{fmfgraph*}
        \end{fmffile}
    }}
    + \vcenter{\hbox{
        \begin{fmffile}{downconversion}
        \begin{fmfgraph*}(50,60)
            \fmfcmd{
                path q; 
                q = (0,0) -- (0.5,0) & quartercircle & (0,0.5) -- (0,0);
                def draw_otimes(expr p) = 
                    filldraw (for i=0 upto 3: q rotated (45+90*i) shifted point 0 of p -- endfor cycle) withpen pencircle scaled 0.4;
                enddef;
            }
            \fmfleft{p1,p2,i1,q1,i2,p3,p4}
            \fmfright{o1}
            \fmfforce{(0.9w,0.2h)}{i1}
            \fmfforce{(0.9w,0.8h)}{i2}
            \fmfforce{(0.15w,0.5h)}{o1}
            \fmfforce{(0.6w,0.5h)}{v1}
            \fmf{photon, arrow, arrow.size=8pt}
            {o1,v1}
            \fmflabel{$\mu$}{o1}
            \fmf{photon, arrow, arrow.size=8pt}{i1,v1}
            \fmflabel{$\beta$}{i1}
            \fmf{photon, arrow, arrow.size=8pt}{i2,v1}
            \fmflabel{$\alpha$}{i2}
            \fmfv{decor.shape=circle, decor.filled=shaded, decor.size=0.16h}{v1}
        \end{fmfgraph*}
        \end{fmffile}
    }}\;\right).
\end{align}
The first term corresponds to SFG (the forward process) while the second term corresponds to SPDC (the reverse process), and $\Pi^\mu_{\alpha\beta}(\Omega;\omega_1,\omega_2)$ is the three-point correlation function, 
\begin{align}
     \Pi^\mu_{\alpha\beta}&(\Omega,\omega_1,\omega_2) = -e^3\int \frac{d^2k}{(2\pi)^2} \sum_{l,m,n}\frac{v^a_{ln}v^a_{nm}v^a_{ml}}{\hbar\Omega-\epsilon_{nl}+i(\eta_l+\eta_n)}\times 
     \\
     &\left(\frac{f(\epsilon_l) - f(\epsilon_m)}{\hbar\omega_1-\epsilon_{ml}+i(\eta_l + \eta_m)}-\frac{f(\epsilon_m) - f(\epsilon_n)}{\hbar\omega_1-\epsilon_{nm}+i(\eta_m + \eta_n)}+\frac{f(\epsilon_l) - f(\epsilon_m)}{\hbar\omega_2-\epsilon_{ml}+i(\eta_l + \eta_m)}-\frac{f(\epsilon_m) - f(\epsilon_n)}{\hbar\omega_2-\epsilon_{nm}+i(\eta_m + \eta_n)}\right),\nonumber
    \label{conducitivity2}
\end{align}
\end{widetext}
The latter expression was obtained using the Keldysh formalism for non-equilibrium Green's functions~\cite{sarsfield2026resonant} aiming to take into account the dominant contributions from nearly resonant intermediate states.

It is important to note that in this formalism, the broadening is attributed to the electronic degrees of freedom, which avoids any spurious divergences and enables us to account for possible band energy dependence of the broadening parameters $\eta_n$.
A brief inspection of Eq. (5) informs us that the optical process involves two types of transitions: one-photon and two-photon processes.
The former corresponds to terms with energy denominators involving $\omega_1$ and $\omega_2$, thus resonating at each photon frequency, while the latter corresponds to those involving $\Omega$, thus resonating at the sum of the two frequencies.
An interesting case appears when both the one-photon and two-photon channels are activated simultaneously, causing a double-resonant condition.
In this case, both the one- and two-photon energy denominators are small, and the total signal will scale as the product of the two, resulting in a strongly enhanced nonlinearity.
This condition is achieved when there are three energy levels for which the two spacings are almost equal to $\hbar\omega_1$ and $\hbar\omega_2$.

\section{Polarization correlations}

Due to in-plane threefold rotational symmetry $C_3$ of all studied graphene multilayers, the tensor components of $\Pi$ are not independent and are related by,
\begin{align}
    \begin{split}
        \Pi^a_{aa} &= -\Pi^z_{az} = -\Pi^z_{za} = -\Pi^a_{zz},
        \\
        \Pi^z_{zz} &= -\Pi^a_{az} = -\Pi^a_{za} = -\Pi^z_{aa}=0.
    \end{split}
\end{align}
The, mirror symmetry with respect to the armchair $a$-axis provides an additional condition that $\Pi^z_{zz}=0$.
Thus, it suffices to compute just one single tensor component, $\Pi^a_{aa}$, to describe nonlinearity involving photons of all possible in-plane polarizations.

\begin{figure*}[hbt!]
\centering
  { \includegraphics[width=1\textwidth]{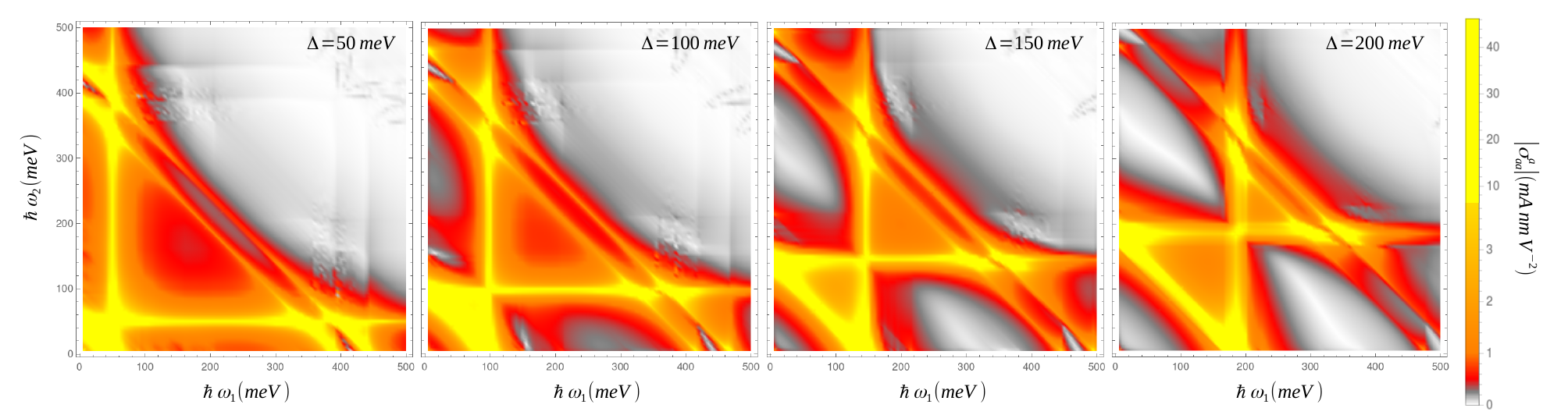}}
\caption{\label{fig:1}
The modulus of second order conductivity for bilayer graphene varying an externally induced symmetry breaking displacement field between the panels. The axis on each panel are the energies of the two lower energy photons respectively.
}
\end{figure*}

To describe SDC and its polarization properties, we calculate the matrix element
\begin{equation}
\sum_{\alpha,\beta}\mel{\omega_1,\alpha;\omega_2,\beta}{H_{\rm int}}{\Omega,\mu}
\end{equation}
For example, for circularly polarised photons, with $\vb*{e}_\pm = (\vb*{e}_x\pm i\vb*{e}_y)/\sqrt{2}$ (where $\mu=\pm$ denotes the helicity), we find,
\begin{widetext}
\begin{align}
    \sum_{\alpha,\beta=\pm}\mel{\omega_1,\alpha;\omega_2,\beta}{H_{\rm int}}{\Omega,\pm}\propto&\sum_{\alpha,\beta}\mel{\omega_1,\alpha;\omega_2,\beta}{\sum_{\mu^\prime,\alpha^\prime,\beta^\prime}\left[\Pi^{\mu^\prime}_{\alpha^\prime\beta^\prime}\,(e^{\mu^\prime}_\Omega)^* e^{\alpha^\prime}_{\omega_1}e^{\beta^\prime}_{\omega_2}\right]^*a^\dagger_{\omega_1,\alpha^\prime}\,a^\dagger_{\omega_2,\beta^\prime}\,a_{\Omega,\mu^\prime}}{\Omega,\pm} \nonumber
    \\
    &=\left[\Pi^\pm_{\mp\mp}\,(e^\pm_\Omega)^*e^\mp_{\omega_1}e^\mp_{\omega_2}\right]^*;\quad \Pi^\pm_{\mp\mp}=\mp i\sqrt{2}\,\Pi^a_{aa}.
    \label{eq:spdc_circular}
\end{align}
\end{widetext}
This suggest that the polarizations of the produced photon pairs $(\alpha,\beta)=(\mp,\mp)$ are reversed with respect to the polarization $\mu=\pm$ of the incoming high-energy photon.
This result is counterintuitive for those accustomed with isotropic media, as it indicates the lack of angular momentum conservation. In fact, the selection rule is a consequence of the $C_3$ symmetry of the crystal, which prescribes quantization of the angular momentum in modulo $3\hbar$. While the incoming photon brings total angular momentum of $\pm\hbar$, the unit of $\pm3\hbar$ is transferred to the crystal, leaving $\mp2\hbar$ to be carried away by the outgoing photon pair.
\begin{comment}
\begin{align}
    \sum_{\alpha,\beta=\pm}\mel{\omega_1,\alpha;\omega_2,\beta}{H_{\rm int}}{\Omega,\pm}\propto\left(\Pi^\pm_{\mp\mp}\,e^\pm_\Omega\,e^\mp_{\omega_1}e^\mp_{\omega_2}\right)^*.
    \label{eq:spdc_circular}
\end{align}
\end{comment}

For linearly in-plane polarized light, with an incoming photon polarized along one of the armchair axes, we find,
\begin{align}
    \sum_{\alpha,\beta=a,z}\mel{\omega_1,\alpha;\omega_2,\beta}{H_{\rm int}}{\Omega,a}& \nonumber
    \\
    &\hspace{-2cm}\propto\left(\Pi^a_{aa}\right)^*\left(e^a_{\omega_1}e^a_{\omega_2}-e^z_{\omega_1}e^z_{\omega_2}\right).
    \label{eq:spdc_lineara}
\end{align}
This suggest that an $a$-polarized photon ($e^a_\Omega=1$) splits into a correlated pair where both outgoing photons have the same linear polarization.

In contrast, for an incoming photon polarized in the $z$-direction ($e^z_\Omega=1$),
\begin{align}
    \sum_{\alpha,\beta=a,z}\mel{\omega_1,\alpha;\omega_2,\beta}{H_{\rm int}}{\Omega,z}& \nonumber
    \\
    &\hspace{-2cm}\propto-\left(\Pi^a_{aa}\right)^*\left(e^a_{\omega_1}e^z_{\omega_2}+e^z_{\omega_1}e^a_{\omega_2}\right),
    \label{eq:spdc_linearz}
\end{align}
which suggest that the two photons in the emitted pair have correlated orthogonal polarizations. 

For an arbitrarily linearly polarized incoming photon with $\vb{e}_\Omega = (\cos\Theta,\,\sin\Theta)$ we find
\begin{align}
    \sum_{\alpha,\beta=\theta_1,\theta_2}\mel{\omega_1,\alpha;\omega_2,\beta}{H_{\rm int}}{\Omega,\Theta}& \nonumber
    \\
    &\hspace{-2cm}\propto\left(\Pi^a_{aa}\right)^*\cos\left(\Theta+\theta_1+\theta_2\right),
    \label{eq:spdc_linear_theta}
\end{align}
where $\vb{e}_{\omega_{1,2}} = (\cos\theta_{1,2},\,\sin\theta_{1,2})$. Equation (10) describes generic correlations between the two generated photons. To mention, the polarization correlations described by the factor of $\cos\left(\Theta+\theta_1+\theta_2\right)$ comply with the $C_3$ symmetry of the crystal. 

For up-conversion the same polarization rules apply to relate the polarizations of photons in the incoming pair to a single higher-energy photon.

\section{Results}
The quantitative analysis using Eq. (5) enables us to characterize up/down-conversion processes in the following multilayer systems: bilayers, Bernal and rhombohedral trilayers and tetralayers of all stable stacking orders. The computed data for up/down-conversion intensity are presented in Fig. 1,2 and 3 in the form of color maps for "second order conductivity", $\sigma^a_{aa}=-\Pi^a_{aa}/(\omega_1\omega_2)$, plotted as a function of the two in/out-going photon frequencies $\omega_1$ and $\omega_2$. To mention, on these maps, the $\omega_1=\omega_2$ line corresponds to the second harmonic generation and the corresponding cut reproduces the earlier published results \cite{sarsfield2026resonant}.

In particular, in Fig. 1, we show the second order conductivity for all stacking orders up to four layer graphene. We notice that while the Bernal stacking orders and the bilayer have little note worthy features the rhombohedral stacking orders posses narrow islands of double resonance around $\hbar\omega_{1,2}\approx400$. Notably the mixed stacking ABCB tetralayer displays several enhanced double resonances.

 To use these maps to discuss up-conversion one can simply identify the spot on the map determined by the incident photon energies to say whether the up-conversion process would be efficient or not. For down-conversion, one should analyze the intensity distribution along the line cuts through the map, at $\Omega=\hbar\omega_1+\hbar\omega_2$. For example, for an ABCB tetralayer, such a line cut at $\Omega\equiv966$meV is presented in the inset, highlighting the double resonance conditions  for the photon splitting into a correlated pair. In this case one may expect the emission to be dominated by [$590$meV,$376$meV] and [$606$meV,$360$meV] pairs.

\begin{figure*}[hbt!]
\centering
  { \includegraphics[width=1\textwidth]{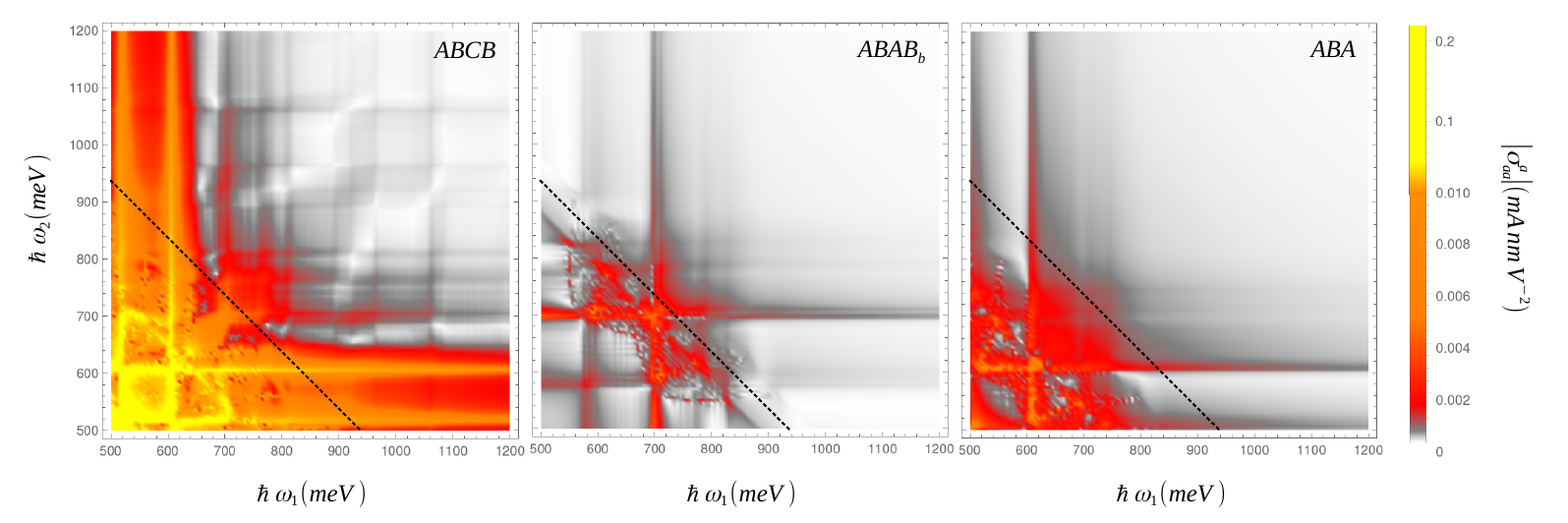}}
\caption{\label{fig:4}
The modulus of second order conductivity for relevant stacking orders up to tetralayer. In this case we show the computed conductivity maps at higher photon energies. The axis on each panel are the energies of the two lower energy photons. The dashed lines indicate energy combinations of the down converted photons which are near the telecoms frequency for optical fibers. 
}
\end{figure*}

In Fig. 2 we show maps for bilayer graphene in the presence of an interlayer asymmetry (e.g. driven by an external displacement field) characterized by parameter $\Delta$.
These maps are instructive to identify double resonance conditions by following the vertical, horizontal and anti-diagonal high intensity lines towards their crossings. The same identification of double resonant features applies to results in Fig. 1 for trilayers and tetralayers. 

Finally in Fig. 3 we present the computed second order conductivity of ABCB, ABAB and ABA in a higher frequency regime. We pick out these stacking orders since they are the only ones displaying noteworthy spectroscopic features in this energy regime. We highlight this regime as it potentially encompasses the telecoms frequency range for the down converted entangled photons ($\hbar\omega_{1,2}\approx800$meV) indicated by the dashed black line. These stacking orders all display well defined resonant and double resonant features at the frequencies that may be utilized for telecoms frequency down conversion. This is particularly convenient for real world applications since large scale growth of graphene multilayers is currently feasible in particular for the bernal stacking configurations \cite{sun2012large}.

\section{Conclusions}
Overall, we have identified few-layer graphene films with up/down-conversion capability in the infrared spectral range and carried out the analysis of polarization correlations, in particular, for down-converted photon pairs. In this analysis, we found the counterintuitive result that for circularly polarized light the polarizations of the produced photon pairs are reversed with respect to the polarization of the incoming high-energy photon. For linearly polarized light we noted the dependence of polarization correlations on the polarization of the incoming high energy photon.

In terms of the appealing spectral interval that would cover the telecom range for either incoming or generated photons, we find that the non-centrosymmetric mixed stacking tetralayer is the most promising candidate. Such ABCB tetralayer poses highly efficient nearly resonant up/down-conversion of [$\omega_1,\omega_2$] photon pairs into/from a photon with $0.7$eV$<\Omega=\omega_1+\omega_2<1.1$eV. We also find pronounced second-order nonlinearity in the spectral range of $0.7$eV$<\Omega<0.9$eV for rhombohedral tri- and tetralayers when asymmetrically encapsulated. In contrast, the Bernal stacking orders only present interesting nonlinearities when looking at higher photon energies as shown in Fig. 3, where we find resonances where down-converted photon pairs are in the telecoms frequency range. 

All this presents the possibility of photon pair production with correlated polarizations generated in optical fibers coated with few-layer graphenes.

\section{Data availability}

The data that supports the findings of this study are available from the corresponding author upon reasonable request.

\begin{acknowledgements}
PJS acknowledges support from CDT Graphene-NOWNANO, and VF from the British Council - ISPF grant 1185409051.
TVJ, TK, and MK were supported by JSPS KAKENHI Grants No. 25K00938, JP24K06921, JP20K14415, No. JP21H05236, No. JP21H05232 by JST CREST Grant No. JPMJCR20T3, and by JST SPRING, Grant No. JP- MJSP2138.
The collaboration between Manchester and Osaka was supported by NPL - ISPF project QEMT.
\end{acknowledgements}

\bibliography{references}
\end{document}